\documentclass[]{spie}  %>>> use for US letter paper
\usepackage[]{graphicx}

\usepackage[colorlinks=False, linkbordercolor={1 1 1}]{hyperref}	

\usepackage[]{times}

\newcommand\arcdeg{\mbox{$^\circ$}}%
\newcommand\ion[2]{#1$\;${\small\rmfamily\@Roman{#2}}\relax}%
\title{Scheduling and calibration strategy for continuous radio monitoring of 1700 sources every three days} 

\author{Walter Max-Moerbeck\supit{a,b}
\skiplinehalf
\supit{a}National Radio Astronomy Observatory, 1003 Lopezville Rd, Socorro, NM, USA\\
\supit{b}California Institute of Technology, 1200 E California Blvd, Pasadena, CA, USA
}

\authorinfo{Further author information: Send correspondence to wmax@nrao.edu \\
Poster presentation at the SPIE Astronomical Telescopes + Instrumentation 2014 in Montr\'eal, Quebec, Canada}
\begin{document} 
  \maketitle

%%%%%%%%%%%%%%%%%%%%%%%%%%%%%%%%%%%%%%%%%%%%%%%%%%%%%%%%
\begin{abstract}
The Owens Valley Radio Observatory 40 meter telescope is currently monitoring a sample of about 1700 blazars every three days at 15 GHz, with the main scientific goal of determining the relation between the variability of blazars at radio and gamma-rays as observed with the \emph{Fermi Gamma-ray Space Telescope}. The time domain relation between radio and gamma-ray emission, in particular its correlation and time lag, can help us determine the location of the high-energy emission site in blazars, a current open question in blazar research. To achieve this goal, continuous observation of a large sample of blazars in a time scale of less than a week is indispensable. Since we only look at bright targets, the time available for target observations is mostly limited by source observability, calibration requirements and slewing of the telescope. Here I describe the implementation of a practical solution to this scheduling, calibration, and slewing time minimization problem. This solution combines ideas from optimization, in particular the traveling salesman problem, with astronomical and instrumental constraints. A heuristic solution using well stablished optimization techniques and astronomical insights particular to this situation, allow us to observe all the sources in the required three days cadence while obtaining reliable calibration of the radio flux densities. Problems of this nature will only be more common in the future and the ideas presented here can be relevant for other observing programs.
\end{abstract}

%>>>> Include a list of keywords after the abstract 


%%%%%%%%%%%%%%%%%%%%%%%%%%%%%%%%%%%%%%%%%%%%%%%%%%%%%%%%
\section{Introduction}

Since mid-2007 we have carried out a dedicated long-term monitoring program of blazars at 15 GHz using the Owens Valley Radio Observatory 40 meter telescope \cite{richards+2011}. The main goal of this program is to study the location of the gamma-ray emission site with respect to the radio emission site by studying correlated variability. The location of the production site of gamma-ray emission in blazars is not well constrained, and different theories and observations favor a location close to the central engine, while others place it at parsec scales in the radio jet. The goal of our program is to reveal if the radio and gamma-ray emission are correlated, which would be a strong indication of a coespatial origin of the photons for these two energy bands. The monitoring of the gamma-ray band is secured by the continuous operation of the \emph{Fermi Gamma-ray Space Telescope} (\emph{Fermi}) which in normal scanning mode scans the whole sky every three hours\cite{atwood+2009}. In order to take full advantage of this capability, a matching program in the radio band is highly desirable, which is the purpose of the OVRO 40 meter program.

Here I describe the OVRO 40 meter blazar monitoring program, as well as the observational techniques used for the radio monitoring. The focus of this paper is on the presentation of the method used to schedule the observation of about 1700 sources every three days, to ensure that we obtain proper calibration and reduce the slewing times to maximize the time the telescope spends doing astronomical observations. The ideas presented here could be of use to other monitoring programs confronting similar challenges.

%%%%%%%%%%%%%%%%%%%%%%%%%%%%%%%%%%%%%%%%%%%%%%%%%%%%%%%%
\section{The OVRO 40 meter telescope radio monitoring program}

\begin{figure}[h!]
\begin{center}
\includegraphics[width=13cm]{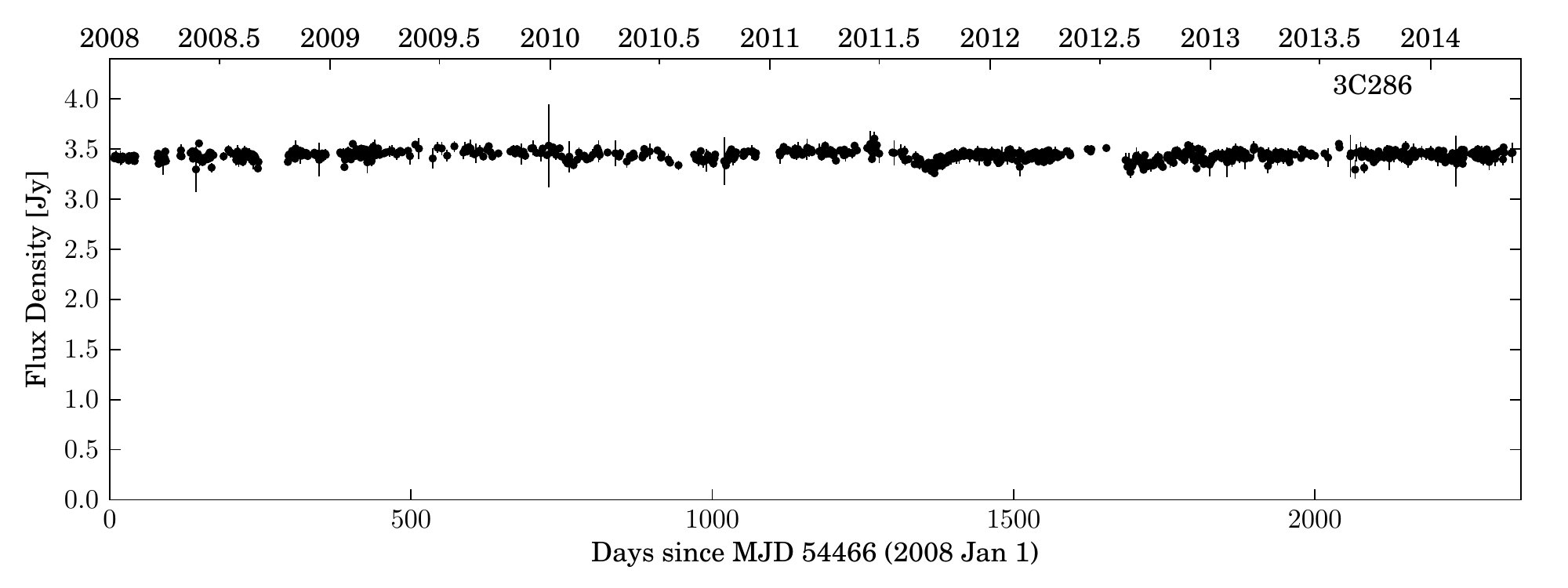}
\includegraphics[width=13cm]{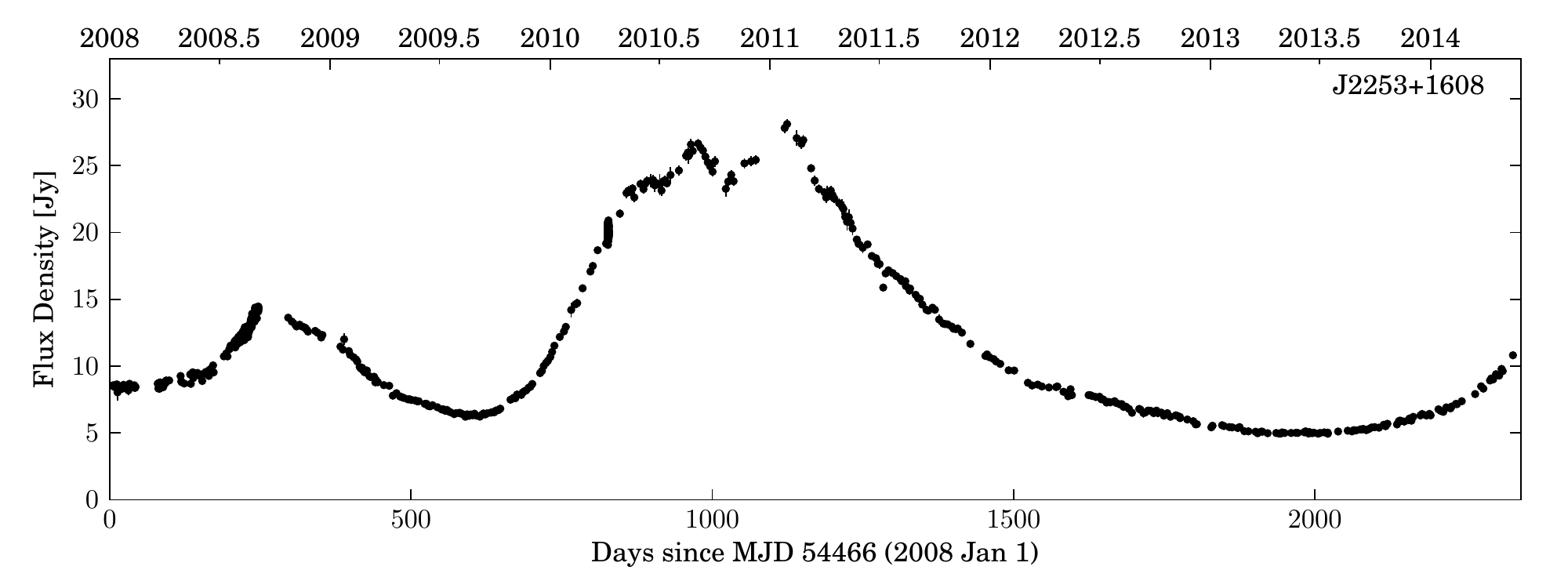}
\caption{Example radio light curves from the OVRO blazar monitoring program. The upper panel is for 3C 286 our primary flux density calibrator. The lower panel is for the blazar J2253+1608, that shows extreme variability in radio and other bands.}
\label{example-radio-light-curve}
\end{center}
\end{figure}

The monitoring program started in mid-2007, before the launch of \emph{Fermi}. The original sample of sources consisted of candidate gamma-ray blazars \cite{healey+2008}, later augmented by adding the blazars detected by \emph{Fermi} which have high confidence associations with radio blazars \cite{abdo+2010, ackermann+2011b}. We observe all the sources in those catalogs located north of $-20\arcdeg$~in declination. Total power flux densities are obtained at 15 GHz with a 3 GHz band, and we attempt to observe each source twice per week. A detailed description of the monitoring program can be found in \cite{richards+2011}. The program generates high quality radio light curves, as example light curves for our primary flux calibrator and the blazar J2253+1608  show in Figure \ref{example-radio-light-curve}. These light curves are the result of continuous monitoring and the scheduling system has played a significant role in making this program possible.

Our OVRO monitoring program has had a broad impact in the community and has produced a large number of publications by our collaboration and other groups.\footnote{More details about our program and its impact can be found in our website \\ \url{http://www.astro.caltech.edu/ovroblazars/}}

% Example of radio light curve for two sources

\section{Scheduling of the observations \label{scheduling_obs}}

The only feasible way to continuously monitor the large number of sources at the cadence required by our program is to make all observations automatic. This demands a scheduling algorithm that can program the observations and calibrations, minimizing the slewing times, and that can run unsupervised. Early in the program we discovered that using the original scheduling and calibration algorithm, some of the sources showed very large flux density variations from day to day, and a subset of them even seemed to have two light curves tracking each other, with one of them at systematically lower flux density levels than the other. The problem was narrowed down to time and sky position variations of the pointing corrections. This section presents a discussion of the requirements and the solution we adopted to solve the scheduling and calibration problem.

\subsection{The problem}

Due to the large aperture of the telescope and the moderately high brightness of our sources, short integrations provide adequate noise levels in most objects. For the monitoring observations we can obtain a thermal noise level of 4 mJy in just 32 seconds of integration, which becomes 63 seconds when we add the overhead associated with moving the telescope between the 4 segments of the beam switched measurement\cite{richards+2011}. A first lower limit in the time required to observe all the sources in the program comes from adding the time it takes to obtain a flux density measurement, which is about $1700 \times 63 / 3600 \approx 30$ hours. Additionally the time needed to move the telescope from source to source can amount to a significant fraction. A quick estimate can be obtained by assuming all the sources are uniformly distributed in the observable region north of $-20\arcdeg$~in declination. This comprises an area of $2.7\pi$ sr, and a mean distance between sources of 4.5\arcdeg. The telescope slews at 15$\arcdeg$ per minute and requires around 12 seconds to settle on each source; taking this into account we can get a lower limit assuming that we can observe all the sources just by jumping into the next one which is at the average distance, in this case we obtain a total slewing time of $1700 \times (12 + 4.5 / (15 / 60))  / 3600 \approx 14$ hours. To these we have to add the time required to calibrate the flux densities and measure pointing model corrections. Relative flux density calibration requires a calibration diode measurement about once an hour, which adds an approximate overhead of one minute every hour of observations, including slewing. Pointing model corrections are the main restriction as we found that these corrections vary with time and are only valid in regions of less than about 25\arcdeg~in diameter. If we divide the sky into about 100 regions, each one requiring a pointing calibration, which takes about 7 minutes, this amounts to about $7 \times 100 / 60 \approx 12$ hours. Adding all these numbers we estimate a lower limit on the total observing time of about 60 hours.

An observing sequence in which we visit each source traveling an average distance of 4.5\arcdeg~is far from what really happens, as it ignores the details of the source distribution and its interaction with other observing constraints such as the limits on zenith angle of the observations. The variation of the pointing model corrections limits the possible source arrangements by forcing us to divide the sky into pointing regions of about 25\arcdeg~in diameter, so the observing problem can be separated into two parts: a first level optimization in which the sources are sorted within an observing region, and a second level optimization in which these regions are sorted. The scheduling problem consists in observing all the sources once in each cycle, minimizing the slewing times and respecting the restrictions imposed by the pointing and calibration requirements, and observability of the sources.

This problem is related to a classical optimization problem, the Traveling Salesman Problem (TSP) in which a traveler salesman has to visit a number of cities minimizing the distance traveled. Each city has to be visited only once and the trip ends in the starting city. This problem does not have a known exact solution and requires numerical techniques to search the solution space and find an appropriate solution, which in most cases is only close to the optimum. A direct search of the solution space is only feasible for small number of cities, making the problem hard even for current computational capabilities. This is due to the fact that the number of possible solutions is $N$!, where $N$ is number of cities, a number that becomes very large even for problems of moderate size (e.g., 100 cities). Various methods have been applied to solve the TSP, among them are the nearest neighbor heuristic, simulated annealing, or problem specific heuristics. The problem has found application in many areas of science and engineering including logistics, genome sequencing, electronic circuit manufacturing and many others. In astronomy it has already been used to help in the scheduling of observations of up to a few hundred sources. Large instances of its most simple form have been solved and software that is able to handle them is freely available. A complete and recent review of solution techniques and references for its various application is given in \cite{applegate+2011}\footnote{A collection of interesting resources related to the TSP can be found in a website maintained by W. J. Cook at \\ \url{http://www.tsp.gatech.edu/}}. Even though the particular example faced by our program is a more complicated version which cannot be handled by these standard tools, some insight can be gained from the existing approaches currently used to tackle this problem.

\subsection{A solution}

In the particular version of the problem our monitoring program needs to solve there are some additional complications. The first one is that the sources are moving in horizontal coordinates, which are the relevant coordinates for the scheduling problem, and the second one is that the sources can only be visited during a time window in which they are sufficiently high in the sky to reduce the effects of the atmosphere and far from the zenith where the telescope has problems tracking the sources. All these factors make the solution more challenging than the traditional TSP problem. After some experimentation with some of the basic tools used for the TSP, a scheme that solves the practical problem of fitting the sources to the desired 3 day cycle was found and is described below.

The requirement of having pointing model corrections for regions of less than $25\arcdeg$ in diameter, requires us to divide the sky into smaller regions. We chose the HEALPix grid to accomplish this as it is widely used in astronomy and has the property of generating regions of equal area and thus a similar number of sources \cite{gorski+2005}. The main restriction of having a distance between the source  and pointing calibrator of less than $25\arcdeg$, can be accomplished by using a grid with 192 pixels over the full sky, each one with $14\arcdeg.7$ of diameter.\footnote{This diameter is actually an equivalent angular size given by $\theta_{\rm pix} = \Omega_{\rm pix}^{1/2}$, where $\Omega_{\rm pix}$ is the solid angle subtended by each pixel.} This requirement reduces the complexity of the optimization problem by naturally dividing the problem into two levels, one in which sources are sorted in a region and a second one in which the regions are sorted to make a complete observing cycle. 

The first step in the optimization is to assign each source to a region. For each region we need to select a source for pointing calibrations. This source is used to obtain the local pointing model corrections that are later applied to the other sources in the region. A pointing calibrator needs to be bright, unresolved and located in a region free of confusion. Almost all the sources in our monitoring program satisfy the unresolved requirement and are far from the Galactic plane in regions which are usually not affected by confusion; the few which are not suitable are excluded from this procedure. The pointing calibrator is chosen from the sources that are brighter than a certain threshold. Regions with one of the standard flux density calibrators use this for pointing. For other regions, we select among the sources brighter than 400 mJy and choose the one with the minimum average distance to other sources. In regions with no sources brighter than 400 mJy we simply choose the brightest source. For a couple of special cases near the Galactic plane where contamination of the reference fields is severe, sources have been incorporated in nearby regions always ensuring a distance of less than 20\arcdeg~to the pointing calibrator.

\begin{figure}[h!]
\begin{center}
\includegraphics[width=10cm]{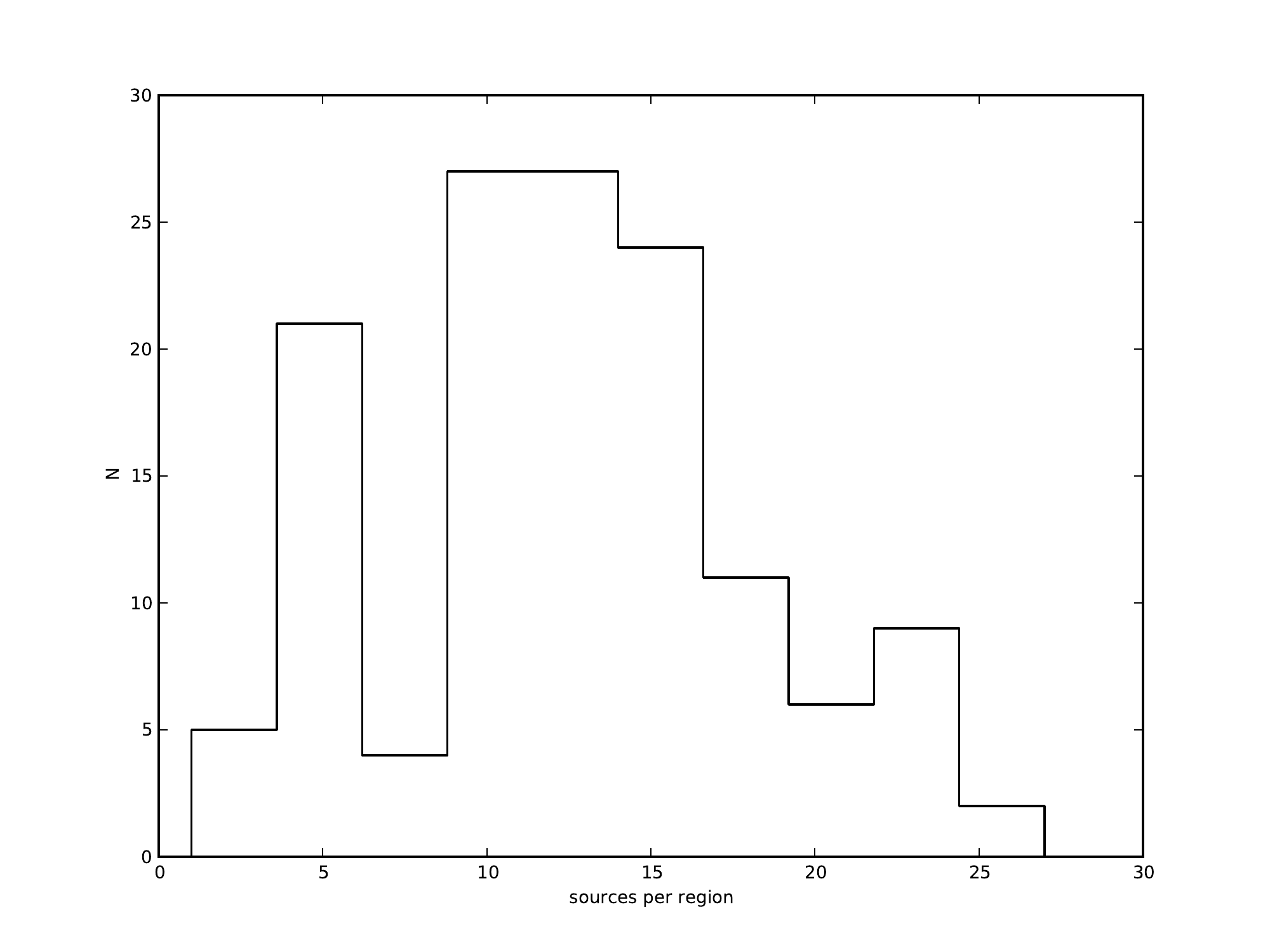}
\caption{Number of sources per region in the sky. Only regions with fewer than 10 sources can be optimized by direct search; for most of them we use simulated annealing.}
\label{number_of_source_per_region}
\end{center}
\end{figure}

For each region we have to solve the problem of visiting the sources starting with the pointing calibrator in an order that minimizes the slewing time. Regions with fewer than 10 sources can be optimized by direct search of the best solution. For regions with at least 10 sources this approach is too costly and we resort to simulated annealing, which can find a good solution in a reasonable time\footnote{The time it takes for a direct search is proportional to the number of possible solutions. Using our implementation on a desktop computer it takes 108 seconds to test the $9!$ possibilities for 9 sources. By simply scaling this we can predict that it would take 300 days for 14 sources and 80\% of the age of the universe for 22 sources. It was not fun to realize this in my second year of graduate school.}. The extent of this problem can be appreciated by looking at Figure  \ref{number_of_source_per_region}, which shows that only a small fraction of the regions can be optimized by direct search.

% Number of sources per region to show the complexity of the problem

Even though we cannot hope to solve the problem exactly with a direct search, we can use simulated annealing, which is an approximation algorithm used to find solutions in optimization problems that do not have exact solutions or when finding such would take an unreasonable amount of time. The basic idea is to simulate the processes of annealing in a metal, in which the material is heated and slowly cooled in order to remove strain and imperfections as a result of minimizing the free energy in the material. In a practical optimization problem, instead of reducing the free energy we are interested in reducing some cost function, and we choose a quantity analogous to the temperature that is a measure of the variability of the cost function. This method is easy to implement and although it does not ensure an optimum solution, it has proven to be efficient at finding good solutions with a reasonable amount of computation \cite{laarhovenAndAarts1987, reinelt1994}. In practice we find appropriate solutions in very short times with only a few thousand iterations. The simulated annealing algorithm is summarized below (adapted from \cite{reinelt1994}). Here a tour is a path that goes from source to source and we denote it by $T$. The time length is the cost function for the minimization and we called it $C(T)$,

\begin{itemize}
\item[\textbullet]
Compute an initial tour of the sources $T$ and choose an initial temperature $\theta > 0$ and a repetition factor $r$.

\begin{itemize}
\item[\textbullet]
As long as the stopping criterion is not satisfied perform the following steps

\begin{itemize}
\item[\textbullet]
Do the following $r$ times.

\begin{itemize}
\item[\textbullet]
Perform a random modification of the current tour $T_{\rm mod}$ and compute the time length difference 
$\Delta = C(T_{\rm mod}) - C(T)$.

\item[\textbullet]
Draw a uniformly distributed random number $x$, $0 \le x  \le 1$.

\item[\textbullet]
If $\Delta < 0$ or $x < \exp(-\Delta / \theta)$ then set $T = T_{\rm mod}$.
\end{itemize}

\item[\textbullet]
Update $\theta$ and $r$
\end{itemize}
\end{itemize}

\item[\textbullet]
Output the current tour as solution

\end{itemize}

This is only a general description of the algorithm that leaves a number of points out. The first one is how to choose a starting temperature $\theta$. There are various recipes for this: one is to explore a number of random paths and use a quantity proportional to the standard deviation on those random tours. In our case we simulate $N$ = 100 random tours and start with $\theta = \theta_i = 3 \sigma$, where $\sigma$ is the variance in those $N$ random tours. Another point is the repetition factor $r$. After some trial we decided to use $r$ = 100, as no significant improvements were found for larger values. The last point is a rule to update $\theta$ and $r$, which is the so-called cooling schedule. We use the simplest cooling schedule in which $r$ is fixed and $\theta$ is reduced by a constant factor, in this case $0.9$. The procedure ends when $\theta < 0.01 \theta_i$. We initialize the optimization by choosing a random tour which is modified at each step by swapping the order of two sources\footnote{In the context of the TSP this is the \emph{city-swap heuristic}}.

\begin{figure}[h!]
\begin{center}
\includegraphics[width=8cm]{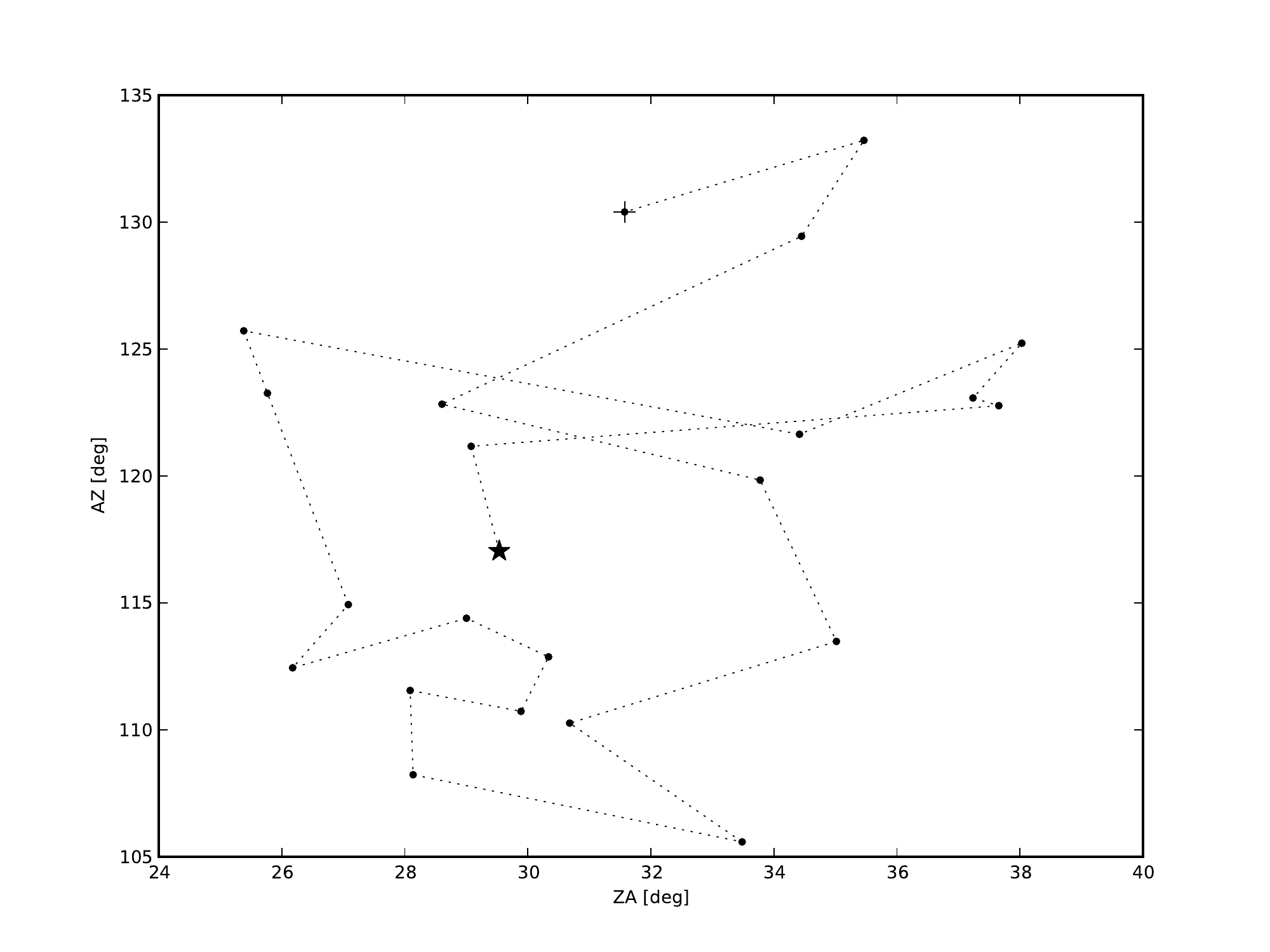}
\includegraphics[width=8cm]{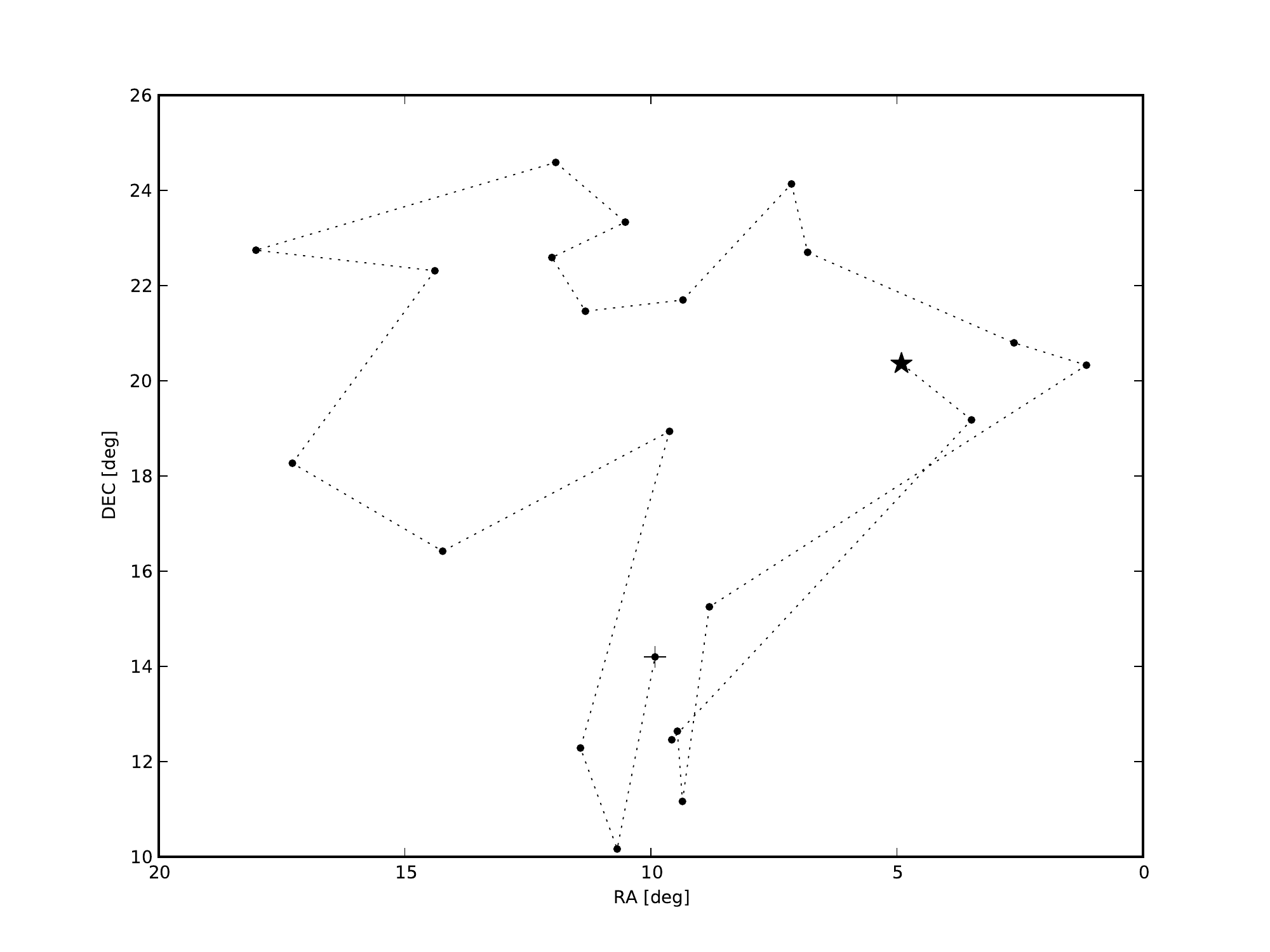}
\caption{Sample path of the telescope for one of the regions in horizontal coordinates (left panel) and equatorial coordinates (right panel). The first source is marked with a star and the last one with a cross. There are a total of 21 sources in this region.}
\label{region_sample_path}
\end{center}
\end{figure}

The solution will depend slightly on the elevation for the observations, but the dependence can be ignored and a fixed value obtained at 45\arcdeg~of elevation can be used for the following stage which is described below. The sample path on the sky for one of the regions is illustrated in Figure \ref{region_sample_path} for a region with 21 sources. 

% Sample path of telescope after optimization

\begin{figure}[h!]
\begin{center}
\includegraphics[width=12cm]{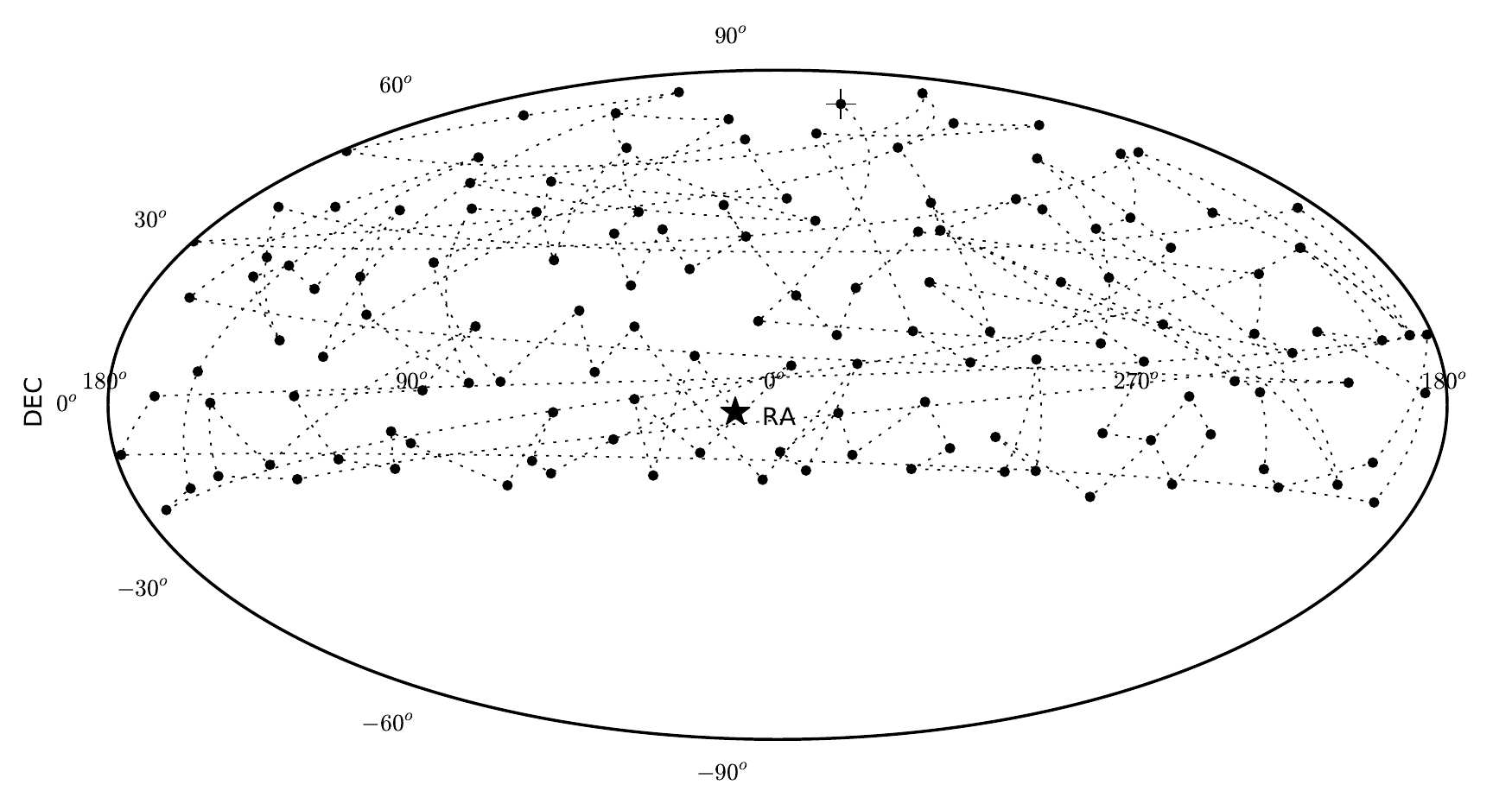}
\caption{Sample telescope path in equatorial coordinates for a full observing cycle of three days. The first region is marked with a star and the last one with a cross.}
\label{sky_sample_path}
\end{center}
\end{figure}

The second stage consists of sorting the regions. This has the extra complication that the regions are only observable for a limited time and that calibrators have to be observed every day. Several approaches were tested, including sorting by declination and right ascension, or by nearest neighbor, but none of them was able to accommodate all the sources in the required 3 day cycle. After some experimentation it was found that a heuristic approach that starts by giving higher priority to observations of southern sources and then moves slightly to the north is able to fit all the sources in the 3 day cycle. This heuristic approach is motivated by the fact that southern sources have a very limited observing window while circumpolar sources are observable at any time. The heuristic approach effectively uses circumpolar sources to fill gaps in the schedule where no other regions are observable. An example path through all the 136 regions with sources is shown in Figure \ref{sky_sample_path}.

% Sample telescope path in RA/DEC for regions in a full observing cycle

This system has been in use since March 2009. It has allowed us to reduce large systematic errors associated with inadequate pointing model corrections for large regions of the sky, a problem that affected a fraction of our early light curves. All those problematic data points have been eliminated and we improved the reliability of our measurements and increased the effective cadence on each source. Although this original heuristic approach has been revised since its adoption we still use the software infrastructure developed in 2009, which has proven to be flexible enough to accommodate new developments.

%%%%%%%%%%%%%%%%%%%%%%%%%%%%%%%%%%%%%%%%%%%%%%%%%%%%%%%%
\section{Summary}

This paper presents the implementation of a practical solution to the telescope scheduling, calibration, and slewing time minimization problem as applied to the OVRO 40 meter telescope blazar monitoring program. This solution combines ideas from optimization with astronomical and instrumental constraints. Problems of this nature are expected to be more common in the future and the ideas presented here can be relevant for other observing programs.

The Owens Valley Radio Observatory 40 meter telescope has been continuously observing blazars since mid-2007. In that time, and enabled in part by the developments presented in this paper, it has been producing high quality light curves used in many refereed publications. Thanks to these data our group has provided insight into the relation between the radio and gamma-ray emission in blazars \cite{ackermann+2011a, pavlidou+2012}, the amplitude of the radio variability and its relation to source class \cite{hovatta+2014, richards+2011, richards+2014}, and the correlations between the radio and gamma-ray variability as a tool to locate the gamma-ray emission site \cite{max-moerbeck+2014}. More than 40 peer reviewed publications, several conference proceeding and Astronomer Telegrams have used our data for multi-wavelength studies of blazars and other variable sources.

%%%%%%%%%%%%%%%%%%%%%%%%%%%%%%%%%%%%%%%%%%%%%%%%%%%%%%%%
\acknowledgments     %>>>> equivalent to \section*{ACKNOWLEDGMENTS}       
 
This work was done while I was a graduate student at Caltech. The successes of the OVRO 40 meter blazar monitoring program are the result of the effort of many people. I thank all the members of the blazar monitoring team at Caltech and elsewhere, for insightful conversations about radio astronomy observations, data analysis techniques and the physics of active galactic nuclei. Particular acknowledgments go to Joseph Richards, Talvikki Hovatta, Anthony Readhead and Tim Pearson, who are the core of the group. I thank Russ Keeney for his support at OVRO. The OVRO program is supported in part by NASA grants NNX08AW31G and NNX11A043G and NSF grants AST-0808050 and AST-1109911. Support from MPIfR for upgrading the OVRO 40-m telescope receiver is acknowledged. The National Radio Astronomy Observatory is a facility of the National Science Foundation operated under cooperative agreement by Associated Universities, Inc.

%%%%%%%%%%%%%%%%%%%%%%%%%%%%%%%%%%%%%%%%%%%%%%%%%%%%%%%%
%%%%% References %%%%%

\bibliography{bibliography}   %>>>> bibliography data in report.bib
\bibliographystyle{spiebib}   %>>>> makes bibtex use spiebib.bst

\end{document}